\documentclass[10pt]{article}
\usepackage{amsmath}
\usepackage{cite} 
\usepackage{indentfirst}
\usepackage{latexsym}
\usepackage{pstricks}

\long\def\ca#1\cb{} 

\newcommand{\ad}{^\dagger }

\newcommand{\becs}{\begin{cases}}
\newcommand{\bem}{\begin{matrix}}

\newcommand{\bsk}{\bigskip }

\newcommand{\dya}[1]{|#1\rangle\langle#1|}
\newcommand{\dyad}[2]{|#1\rangle\langle#2|}

\newcommand{\encs}{\end{cases}}
\newcommand{\enm}{\end{matrix}}

\newcommand{\hf}{{\textstyle\frac{1}{2} }}

\newcommand{\ket}[1]{|#1\rangle }

\newcommand{\lra}{\leftrightarrow }

\newcommand{\msk}{\medskip }

\newcommand{\od}{\odot }

\newcommand{\ra}{\rightarrow }
\newcommand{\Ra}{\Rightarrow }

\newcommand{\st}{\sqrt{2}}

\newcommand{\vb}{\,|\,}

\newcommand{\HC}{{\mathcal H}}

\newcommand{\al}{\alpha }

\def\outl#1{\par{\medskip\noindent\hspace*{.5cm}\bf
      \mathversion{bold}#1\mathversion{normal}\smallskip} }
  \def\xa{} \def\xb{}  

 \def\outl#1{}   \def\xa{} \def\xb{}  

\ca
 \def\outl#1{\par{\medskip\noindent\hspace*{.5cm}\bf
      \mathversion{bold}#1\mathversion{normal}\smallskip} }
 \long\def\xa#1\xb{}
 
\cb

\begin{document}

\ca
\centerline{\Large Quantum Measurements Are Noncontextual}
\xa
\msk
\cb

\title{Quantum Measurements are Noncontextual}

\author{Robert B. Griffiths
\thanks{Electronic mail: rgrif@cmu.edu}\\ 
Department of Physics,
Carnegie-Mellon University,\\
Pittsburgh, PA 15213, USA}
\date{Version of 24 April 2013}
\maketitle  

\xb

\xa
\begin{abstract}

  Quantum measurements are noncontextual, with outcomes independent of which
  other commuting observables are measured at the same time, when consistently
  analyzed using principles of Hilbert space quantum mechanics rather than
  classical hidden variables.
\end{abstract}
\xb

\bsk
\xa

\xb \outl{Bell: Issue is QM is contextual? raised in discussion of hidden
  variables} 
\xa

John Bell in Sec.~5 of \cite{Bll66} while discussing hidden variables in
quantum mechanics raised the question of whether quantum theory is
``contextual.''  In particular, does the outcome of a measurement of an
observable $A$ on a system $S$ depend on whether $A$ is measured along with an
observable $B$ that commutes with $A$, or along with a different observable $C$
that also commutes with $A$ but does not commute with $B$?  (In what follows we
make no distinction between an observable $A$ and the operator $A=A\ad$ that
represents it on the Hilbert space $\HC_S$ of the system in question.)
Everyone agrees that if two observables commute they can be measured
simultaneously, so there is no difficulty imagining $A$ and $B$ measured
simultaneously on a single system using a single apparatus. Similarly one can
imagine $A$ and $C$ measured simultaneously by a single apparatus, but since
$B$ and $C$ do not commute, the two apparatuses just mentioned must be
different. Suppose the first apparatus yields some value of $A$.  Would the
second have yielded the same result?  A theory which answers ``Yes'' is said to
be \emph{noncontextual}; one which answers ``No'' is \emph{contextual}, i.e.,
the measurement outcome depends on the context, on what else is being measured
at the same time.  Note that the issue being addressed here is that of
\emph{individual} measurement outcomes, not the probability of an outcome in a
situation where measurements may be repeated a large number of times. 

\xb
\outl{Will argue that QM is noncontextual if apparatus and analysis is fully
  Qml}
\xa

\xb
\outl{Techniques used for this analysis have not been mentioned in recent
  papers}
\xa

It will be argued below that quantum theory is \emph{noncontextual}: the
measured value of $A$ does \emph{not} depend on which other observable,
necessarily one commuting with $A$, is measured simultaneously with it,
assuming the measurement is carried out with properly designed apparatus.  The
basic point is that the outcome of such a measurement can be interpreted as
revealing a property that the measured system---we refer to it as a
\emph{particle}---had \emph{before} the measurement took place. By a quantum
property we mean (following von Neumann, see Sec.~III.5 of \cite{vNmn32b})
something represented by a subspace of the appropriate quantum Hilbert space,
equivalently the projector onto this subspace, rather than by some additional
hidden variable that is not part of the Hilbert space. The techniques needed
for this analysis have been available for well over a decade
\cite{Grff84,Grff02c}, but seem to have been ignored in a substantial
collection of papers---the number is large, and what follows is far from an
exhaustive list---published relatively recently in this
\cite{CbCn11,VdWh11,Amao12,KrRK12,Zuao12,Klao12,Cbll13,Amao13,Zuao13,TnYO13}
and in other journals \cite{Spkk05,Appl05b,Gnvs05,Cbll11,Lpao11,LnSW11,BFGO11,
  Hsgw12,Cbll12,Cbao12,AbHr12,KrKs12,Cbll13b,Chao13}.  (Some older discussions
of the contextuality problem in \cite{Mrmn93} and in Ch.~7 of \cite{Prs93} are
in certain respects clearer than Bell's original work. The philosopher's
perspective is well represented in \cite{Hld06} and \cite{Hrmn11}.) These
papers leave the reader with the misleading (in our opinion) impression that
quantum mechanics is contextual because it fails to satisfy certain
inequalities derived on the basis of classical hidden variable theories. We
believe that, on the contrary, the real issue is that quantum mechanics is
\emph{not} classical mechanics, and when analyzed using conceptual tools
consistent with its mathematical (Hilbert space) structure quantum mechanics is
noncontextual, using this term in the sense originally employed by Bell.

\xb
\outl{Two measurement problems not resolved by Bell or by textbooks}
\xa

Textbook quantum theory cannot address Bell's question for reasons Bell himself
pointed out in one of his last publications \cite{Bll90}: it employs
``measurement'' as a sort of fundamental principle or axiom, a black box which
cannot be opened or further analyzed.  This is the great ``measurement
problem'' of quantum foundations, which in fact is two problems.  The first is
that the unitary time development produced by Schr\"odinger's equation when
applied to the apparatus as well as the system being measured can lead to a
superposition of outputs---pointer positions in the quaint but picturesque
language of quantum foundations---which is hard to interpret.  Disposing of
this ``Schr\"odinger cat,'' or, speaking metaphorically, stopping the pointer
from wiggling, constitutes the \emph{first} measurement problem.  But when this
has been solved the \emph{second} measurement problem remains: inferring from
the macroscopic pointer position some microscopic state of affairs that existed
\emph{before}, not after, the measurement took place. Such inferences are
frequently carried out by experimental physicists: think of a neutrino coming
from the sun, or a gamma ray emerging from a nucleus and gobbled up by a
detector.  The widespread idea that a quantum measurement only tells one
something about what exists \emph{after} the measurement is complete arises
from an inadequate treatment in textbooks, in which a \emph{measurement} is
often confused with the \emph{preparation} of a system in a particular quantum
state.  The two are not unrelated, but they are in fact distinct; see
\cite{Grff13} for further discussions of this point.

\xb
\outl{CH approach is the only Qm interpretation answering two measurement 
problems}
\xa

\xb
\outl{Previous literature on CH}
\xa

\xb
\outl{Following discussion accessible to readers who don't know CH}
\xa

The consistent or decoherent histories approach, hereafter referred to as
``histories,'' seems at present the only interpretation of quantum mechanics
employing the quantum Hilbert space without using additional hidden variables
that gives satisfactory answers to both measurement problems. It is the basis
of the analysis that follows. There are by now numerous expositions of the
basic histories approach; in order of decreasing length we recommend the
following: \cite{Grff02c,Grff11b,Hhnb10,Grff09b}.  There have been numerous
criticisms; for an analysis of, and response to the major ones the reader is
referred to \cite{Grff13} and other work cited there. Its proponents do not
regard the histories approach as antithetical to standard quantum mechanics
as found in textbooks.  It predicts exactly the same probabilities for
measurement outcomes, but in addition it allows measurements themselves to be
analyzed in fully quantum mechanical terms, and in particular the outcomes
of measurements to be related to the microscopic properties the apparatus was
designed to measure. The following discussion should for the most part be
accessible to readers familiar only with the standard (textbook) approach,
though at certain points reference is made to published results in order to
shorten a longer discussion.  A more detailed but also more abstract and less
physical approach to the issue of contextuality will be found in
\cite{Grff13b}.


\xb
\outl{Example for 3-dim Hilbert space.  Observables A, B, C defined}
\xa

Rather than discussing abstract principles, let us consider a specific example
in which $\HC_S$ is three dimensional (think of the spin of a spin-one
particle, though angular momentum does not enter the following discussion),
with an orthonormal basis $\ket{1},\,\ket{2},\,\ket{3}$, and three different
observables defined as follows using dyads:
\begin{equation}
  A = \dya{1} - \dya{2}-\dya{3},\quad
 B = \hf\dya{1} +\dya{2} -\dya{3},\quad
 C = 2\dya{1} +\dyad{2}{3} + \dyad{3}{2}.
\label{eqn1}
\end{equation}
It is obvious that $[A,B] = 0$, and straightforward to show that $[A,C] = 0$
and $[B,C]\neq 0$.

\begin{figure}[h]

\xb
\outl{Figure 1 showing apparatus}
\xa

$$
\begin{pspicture}(-1,-1.5)(8.2,1.5) 
\newpsobject{showgrid}{psgrid}{subgriddiv=1,griddots=10,gridlabels=6pt}
\def\lwd{0.035} 
\def\lwb{0.10}  
\def\lwn{0.01}  
\psset{
labelsep=2.0,
arrowsize=0.150 1,linewidth=\lwd}
  \def\rarr(#1){\rput(#1){\psline{->}(-0.2,0)(0,0)}} 
  \def\larr(#1){\rput(#1){\psline{->}(0.2,0)(0,0)}}  
  \def\uarr(#1){\rput(#1){\psline{->}(0,-0.2)(0,0)}} 
  \def\darr(#1){\rput(#1){\psline{->}(0,0.2)(0,0)}}  
  \def\ruarr(#1){\rput(#1){\psline{->}(-0.15,-0.15)(0,0)}}  
  \def\rdarr(#1){\rput(#1){\psline{->}(-0.15,+0.15)(0,0)}}  
  \def\luarr(#1){\rput(#1){\psline{->}(+0.15,-0.15)(0,0)}}  
  \def\ldarr(#1){\rput(#1){\psline{->}(+0.15,+0.15)(0,0)}}  
\def\dput(#1)#2#3{\rput(#1){#2}\rput(#1){#3}}
\def\rectg(#1,#2,#3,#4){
\psframe[fillcolor=white,fillstyle=solid](#1,#2)(#3,#4)}
\def\hdg{0.5} \def\squ{%
\psframe[fillcolor=white,fillstyle=solid](-\hdg,-\hdg)(\hdg,\hdg)}
\def\drad{0.35}\def\dradp{0.247}
\def\detectur{
\psarc[fillcolor=white,fillstyle=solid](0,0){\drad}{-45}{135}
\psline(-\dradp,\dradp)(\dradp,-\dradp) }
\def\detectdr{
\psarc[fillcolor=white,fillstyle=solid](0,0){\drad}{225}{45}
\psline(-\dradp,-\dradp)(\dradp,\dradp) }
\psline{>-}(-1,0.0)(6,0.0)
\psline(1.3,0)(2.3,1)
\rectg(0,-0.5,1.6,0.5)
 
\psline(6.3,0)(7.3,1)
\psline(6.3,0)(7.3,-1)
\rectg(5,-0.5,6.6,0.5)
\ruarr(2,0.7)
\ruarr(7,0.7)
\rdarr(7,-0.7)
\rarr(2.2,0)
\rarr(4.8,0)
\rput(2.3,1){\detectur}
\rput(7.3,1){\detectur}
\rput(7.3,-1){\detectdr}
\psline(2.5,0.2)(3,0.2)
\psline(2.5,-0.2)(3,-0.2)
\dput(3.9,0){\squ}{$U$} 
\rput[t](2.75,-.3){$M_1$}
\rput[l](2.7,1.1){$D_1$}
\rput[l](7.7,1.1){$D_2$}
\rput[l](7.7,-1.2){$D_3$}
\rput(0.75,0){$V$}
\rput(5.75,0){$W$}
\end{pspicture}
$$
\caption{Apparatus to measure $A$ along with $B$ ($U=U_B$), or with $C$ 
($U=U_C$).}
\label{fgr1}
\end{figure}
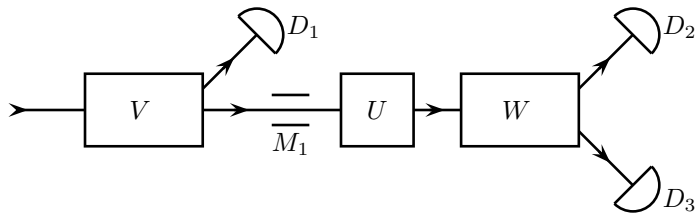

\xb
\outl{Description of measurement process}
\xa

An apparatus for measuring these observables is shown schematically in
Fig.~\ref{fgr1}.  The incoming particle first passes through a device $V$ (one
can think of an electric field gradient acting on a particle with an electric
quadrupole moment) which splits the path in two: the upper path, followed by a
particle in the state $\ket{1}$, leads to the detector $D_1$. The lower
(straight) path, followed by a particle whose state is any linear combination
of $\ket{2}$ and $\ket{3}$, passes through a nondestructive detector $M_1$ that
measures the particle's passage without disturbing its internal state, and then
through another device $U$ that carries out a unitary transformation $U_B$
equal to the identity $I$ (i.e, the device does nothing) if $B$ is to be
measured, or
\begin{equation}
 U_C = (1/\st)\bigl\{\dya{2} + \dyad{2}{3} + \dyad{3}{2}  -\dya{3}\bigr\}.
\label{eqn2}
\end{equation}
if $C$ is to be measured.  Following this yet another device $W$ (e.g., think
of a Stern-Gerlach magnet) splits the trajectory into one moving upwards if the
particle state is in state $\ket{2}$, or downwards if it is in state $\ket{3}$;
these terminate in detectors $D_2$ and $D_3$.

\xb
\outl{What happens if incoming particle is in eigenstate of A or of B or of C}
\xa

A particle initially in the eigenstate $\ket{1}$ of $A$ with eigenvalue $+1$
will be detected by $D_1$, whereas any eigenstate with eigenvalue $-1$, some
linear combination of $\ket{2}$ and $\ket{3}$, will be detected by $M_1$ and
then travel on. Thus a measurement of $A$ precedes the particle's passing
through the box $U$, and the outcome will not be affected by whether the
unitary is $U_B$ or $U_C$. Which of these is present could, in principle, be
decided at the very last moment, after the particle (if on this path) has
passed through $M_1$. A measurement of $B$ is carried out by setting $U=I$, so
that initial eigenstates with eigenvalues of 1/2, 1, and $-1$ will be detected
by detectors 1, 2, and 3, respectively.  Alternatively, $C$ can be measured by
setting $U=U_C$, \eqref{eqn2}; eigenvalues of 2, 1, and $-1$ correspond to
detection by detectors 1, 2, and 3, respectively.

\xb
\outl{If initial state is not eigenstate use Born to get rid of Schr cat}
\xa

But suppose the incoming particle has been prepared in a state $\ket{\psi_0}$
which is \emph{not} an eigenstate of the operators which will later be
measured.  How can one avoid Schr\"odinger's cat, if the detectors themselves
are quantum devices, as assumed by most physicists nowadays?  The cleanest way
to resolve this (first) measurement problem is to employ Born's idea
\cite{Brn26} that a quantum wave function evolving unitarily in time is not to
be regarded as physical reality, but instead interpreted using
probabilities---the wave function is a ``pre-probability'' in the notation of
Ch.~9 of \cite{Grff02c}.  The modern approach is to set up a suitable
framework, a collection of quantum histories in which the ordinary macroscopic
outcome results for the detectors are represented by appropriate
(``quasi-classical'') projectors on the full quantum Hilbert space of
particle-plus-measuring apparatus.  For details, see \cite{Grff02c}, in
particular Chs.~17 and 18. 

\xb
\outl{If future does not influence past, later $B\ra C$ change cannot affect
  $A$ outcome }
\xa

\xb
\outl{For general case, need to address second measurement problem}
\xa

It is, of course, clear given the construction shown in Fig.~\ref{fgr1} that if
the change from an $A$-plus-$B$ apparatus to an $A$-plus-$C$ apparatus is made
after the particle has passed the position of detectors $D_1$ and $M_1$, this
cannot affect the $A$ measurement outcome, at least if the future does not
influence the past, so in this sense it seems clear that this measurement is
noncontextual.%
\footnote{A referee raised the issue of whether the future not influencing the
  past could be part of a ``classical intuition'' inconsistent with quantum
  mechanics.  There is no evidence to suggest that histories quantum mechanics
  violates the second law of thermodynamics, though the subject needs further
  investigation; see Sec.~8.4.3 of \cite{Grff13}. Even in classical statistical
  mechanics there is no really satisfactory derivation of the second law from
  microscopic dynamics, so no precise proof that the future does not influence
  the past.}%
However, this might leave open the possibility that in some other measurement
setup the measurement of $C$ instead of $B$ would affect the $A$ measurement
outcome.  In order to dispose of this concern we need to address the second
measurement problem and show that the outcome of the $A$ measurement reflects a
property possessed by the particle before the measurement took place, and this
is true (for a properly constructed apparatus) whatever state $\ket{\psi_0}$
the particle is initially prepared in.

\xb
\outl{Histories that include properties of $A$ just before measurement}
\xa

Let $t_1$ be a time just before the particle reaches the measuring device,
e.g., before it enters the $V$ box in Fig.~\ref{fgr1}, and
assume that its unitary time evolution (it is traveling in a field-free region)
from $t_0$, when initially prepared, to $t_1$ is trivial: $\ket{\psi_1} =
\ket{\psi_0}$. At time $t_1$ introduce a projective decomposition of the
identity on $\HC_S$,
\begin{equation}
 I = \dya{1} + \Bigl(\dya{2} + \dya{3}\Bigr) = P_1 + P_2
\label{eqn3}
\end{equation}
and consider a family of four histories at times $t_0 < t_1 < t_2$,
\begin{equation}
 [\Psi_0] \od \{P_1,P_2\} \od \{D_1, M_1\}.
\label{eqn4}
\end{equation}
Here $[\Psi_0] = \dya{\Psi_0}$ is the projector corresponding to a state
$\ket{\Psi_0}$ at $t_0$ which is a tensor product of an initial apparatus state
with the particle state $\ket{\psi_0}$;%
\footnote{The preparation apparatus could also be included, but this adds
  nothing but a slight complication to the following discussion} %
$P_1$ and $P_2$ are projectors on the particle properties defined in
\eqref{eqn3}, understood to be tensored with the identity operator on the
apparatus Hilbert space; and at a time $t_2$, after the measurement is
complete, $D_1$ and $M_1$ are projectors corresponding to the two macroscopic
outcomes.  Here $\od$ is a tensor product symbol, but for present purposes it
can simply be regarded as separating events at different times. The first of
the four histories, $[\Psi_0] \od P_1 \od D_1$, has the physical interpretation
that the particle was in state $\ket{1}$ at time $t_1$, and at time $t_2$
detector $D_1$ has registered its arrival.  The other three possibilities, with
$P_2$ replacing $P_1$ or $M_1$ replacing $D_1$, are interpreted in a similar
way.

\xb
\outl{Analysis to show particle had property of $A$ corresponding to
  measurement outcome}
\xa

A relatively straightforward probabilistic analysis---for details, see the
examples in \cite{Grff11b} or Chs.~17 and 18 of \cite{Grff02c}---of this family
of histories yields, in the case of a properly constructed measurement
apparatus, conditional probabilities
\begin{equation}
 \Pr(P_1\vb D_1)=1;\quad \Pr(P_2\vb M_1)=1.
\label{eqn5}
\end{equation}
In words, if at time $t_2$ detector $D_1$ has detected the particle, then one
is certain (probability 1) that at time $t_1$ the particle had the property
$P_1$, i.e., the value of $A$ was $+1$, whereas if the particle's passage was
indicated by $M_1$ then it is certain that at time $t_1$ it had the property
$P_2$ and the value of $A$ was $-1$.  One can regard \eqref{eqn5} as
constituting an essential feature of what one means by a competently designed
and built apparatus for measuring $A$, whether or not it takes the form shown
schematically in Fig.~\ref{fgr1}: the macroscopic outcome must be able to
reveal the prior microscopic state.

\xb
\outl{Textbook analysis cannot address second measurement problem}
\xa

By contrast, the assumption made implicitly in textbooks is that in
place of $P_1$ and $P_2$ one should at time $t_1$ use the projective
decomposition\begin{equation}
 I = \dya{\psi_1} + \bigl( I - \dya{\psi_1}\bigr) = Q_1 + Q_2,
\label{eqn6}
\end{equation}
where $Q_1$ corresponds to unitary time development of the particle state and
$Q_2$ to its negation. There is nothing wrong with this choice. However, in
general $Q_1$ and $Q_2$ will not commute with the projectors $P_1$ and $P_2$
needed to discuss the microscopic properties the apparatus was designed to
measure.  There is no way of consistently combining noncommuting projectors in
a meaningful quantum description; pretending that this is possible leads to
paradoxes.  The choice between $\{Q_1,Q_2\}$ and $\{P_1,P_2\}$ at the
intermediate time depends on whether one wishes to relate the properties of the
particle at this time to its earlier preparation or to the outcome of a later
measurement.%
\footnote{A referee has asked what happens if at a later time $Q_1$ and $Q_2$
  are measured rather than $P_1$ and $P_2$.  If the $Q_j$ and $P_j$ do not
  commute this cannot be done in a single experimental run.  One can ask the
  \emph{counterfactual} question of what \emph{would have} happened if the
  $Q_j$ had been measured instead of the $P_j$. Quantum counterfactuals are
  discussed in detail in Ch.~19 of \cite{Grff02c} and applied to Hardy's
  paradox in Ch.~25.  Again, there is no indication that the future influences
  the past.} %
This does not at all imply that the property at the time $t_1$ depends in any
causal sense on which future measurement the particle might undergo; for more
on this topic see Sec.~14.4 of \cite{Grff02c} (where, the reader should be
warned, the term `contextual' is used in a sense different from that employed
by Bell and used in the present Letter).  For additional discussion of the
issue of incompatible frameworks and why they cannot be combined (the single
framework rule) see \cite{Grff13}, and for specific examples see Sec.~4.6 of
\cite{Grff02c}, and \cite{Grff11b}.

\xb
\outl{Generalization from 3 to arbitrary number of states: PD for $A$}
\xa

The preceding discussion for a particle with three states is easily
generalized to the case of an arbitrary (finite) number of states.  To see
this, let $\{P_\al\}$ be the projective decomposition which diagonalizes $A$ in
the sense that the eigenvalues in
\begin{equation}
 A = \sum_\al a_\al P_\al
\label{eqn7}
\end{equation}
are distinct: $a_\al\neq a_{\al'}$ when $\al\neq\al'$. Then it is
straightforward to show that if $A$ commutes with $B$, every $P_\al$ in
\eqref{eqn7} also commutes with $B$.  That is, if a basis is chosen such that
the matrix of $A$ is diagonal with identical eigenvalues placed in separate
blocks, the corresponding matrix of $B$ is block diagonal, and each of its
blocks can be separately diagonalized by a change of basis that leaves the $A$
matrix unchanged.  The same comment applies to any other
observable $C$ that commutes with $A$, whether or not it commutes with $B$,
though of course the bases used to diagonalize $B$ and to diagonalize $C$ must
be different if $[B,C]\neq 0$.

\xb
\outl{How to construct measuring apparatus for an arbitrary number of states}
\xa

A measuring apparatus similar to that in Fig.~\ref{fgr1} can be constructed by
first separating the incoming particle trajectory into different paths
corresponding to the different $P_\al$ (thus different $a_\al$ values) in
\eqref{eqn7}, and then using nondestructive measurement devices (like $M_1$ in
Fig.~\ref{fgr1}) to detect which path the particle is moving along without
affecting its internal state.  Following this, on each path install an
appropriate unitary, which will in general be different depending on whether
one wishes to measure $B$ or $C$, and after each unitary a device to produce a
further separation of paths directed into the final detectors. Since the $A$
measurement can be made in advance of either $B$ or $C$, it is evident that
this apparatus design provides a noncontextual measurement.  A histories
analysis analogous to that given in \eqref{eqn3} to \eqref{eqn5} then shows
that this result is completely general and applies to \emph{any} apparatus
properly designed and constructed to accurately measure $A$ together with (or
without) any other observable that commutes with $A$.  The measurement is
noncontextual, for the outcome reveals a property the particle possessed before
the measurement began.

\xb
\outl{Prior property $\lra$ measurement outcome $\Ra$ QT is noncontextual}
\xa

\xb
\outl{Why the common claim that Qm measurements are \emph{contextual}?\\
1. Measurements not properly treated in textbooks\\
2, Hidden variables are classical, not quantum, objects}
\xa

\xb
\outl{}
\xa

It is this link between a measurement outcome and a property of the measured
system before the measurement took place that demonstrates that in quantum
theory the measurement process is noncontextual.  So why is it that one so
often hears the contrary?  A number of reasons come to mind.  First,
measurements are inadequately treated in textbooks; one admires authors (e.g.,
\cite{Lloe01,Lloe12}) who are brave enough to concur publicly with Bell
\cite{Bll90}: here is a problem they have not been able to resolve.  But of
course without some, at least implicit, theory of quantum measurements one
cannot even begin to discuss contextuality.  Second, in attempting to fill this
serious gap in standard (textbook) quantum mechanics it has been assumed by
Bell and many others that microscopic properties might be represented not by
Hilbert subspaces, but by hidden variables which are in certain essential
respects \emph{classical}. This is obvious in the best-known hidden variable
approach, the de Broglie-Bohm pilot wave \cite{BcVl09,Glds12}, where the
particle is assumed to have a well-defined position at all times. The hidden
variables approach seems almost inevitably to lead to the conclusion that
quantum mechanics is infested with nonlocal influences, and the choice of a
measurement at some spacelike separated point can influence what is going on
here, a ``contextual'' influence.
If one uses the histories approach based on a proper Hilbert space analysis
with no hidden variables such nonlocality disappears; see the discussion in
\cite{Grff11}, and in a more pedagogical form in \cite{Grff11b}, and with it
any notion that violations of Bell inequalities support the idea of quantum
contextuality.  

Third, the Bell-Kochen-Specker (BKS) result, see e.g. \cite{Mrmn93}, is often
invoked as grounds for contextuality.  For details of the analysis showing that
BKS does not imply contextuality we refer the reader to Ch.~22 in
\cite{Grff02c}, to the discussion in Sec.~5 of \cite{Grff13b} referring to the
concept of ``realism'' as presented in \cite{Hrmn11}, and to earlier work in
\cite{Grff00b}. To put the matter briefly, BKS is perfectly fine as a
mathematical theorem.  However, the collections of Hilbert-space projectors
used to construct a BKS paradox do not commute with each other, and hence
cannot all be used together in a single consistent quantum description of a
physical state of affairs; in histories terminology the single framework rule
is violated.  What BKS is telling us is not that quantum measurements are
contextual, but that the textbook treatment of measurements is inadequate for
understanding the quantum world.

In summary, decades of research have shown that trying to understand quantum
mechanics using hidden variables always leads sooner or later to serious
conceptual problems such as a peculiar action-at-a-distance inconsistent with
special relativity. Or to measurement contextuality, which, were it true, would
seriously undermine confidence in experimental results.  By contrast, a
consistent Hilbert space approach, with quantum properties represented by
Hilbert subspaces, is internally consistent, paradox free (so far as we know at
present), and applies without exception to systems of arbitrary size and
complexity.  Students are taught that they cannot always write $XY=YX$ in
quantum physics, even in cases where this is perfectly fine in classical
physics; one must pay attention to whether operators commute. They need to
learn the corresponding rules for reasoning consistently about quantum
properties, and not simply be told ``Shut up and calculate; quantum mechanics
only predicts outcomes of measurements.''  They don't believe it, and they
shouldn't.

\section*{Acknowledgments}

I am grateful to an anonymous referee for comments leading to a clearer
presentation. This research received financial support from the National
Science Foundation through Grant PHY-1068331.

\xa

\xb

\begin{thebibliography}{10}

\bibitem{Bll66}
John~S. Bell.
\newblock On the problem of hidden variables in quantum mechanics.
\newblock {\em Rev. Mod. Phys.}, 38:447--452, 1966.
\newblock reprinted in \cite{Bll87}, pp.~1-13.

\bibitem{vNmn32b}
Johann von Neumann.
\newblock {\em Mathematische Grundlagen der Quantenmechanik}.
\newblock Springer-Verlag, Berlin, 1932.
\newblock English translation: \textit{Mathematical Foundations of Quantum
  Mechanics}, Princeton University Press, Princeton (1955).

\bibitem{Grff84}
Robert~B. Griffiths.
\newblock Consistent histories and the interpretation of quantum mechanics.
\newblock {\em J. Stat. Phys.}, 36:219--272, 1984.

\bibitem{Grff02c}
Robert~B. Griffiths.
\newblock {\em Consistent Quantum Theory}.
\newblock Cambridge University Press, Cambridge, U.K., 2002.
\newblock http://quantum.phys.cmu.edu/CQT/.

\bibitem{CbCn11}
Ad\'an Cabello and Marcelo~Terra Cunha.
\newblock Proposal of a two-qutrit contextuality test free of the finite
  precision and compatibility loopholes.
\newblock {\em Phys. Rev. Lett.}, 106:190401, 2011.

\bibitem{VdWh11}
Thomas Vidick and Stephanie Wehner.
\newblock Does ignorance of the whole imply ignorance of the parts? {L}arge
  violations of noncontextuality in quantum theory.
\newblock {\em Phys. Rev. Lett.}, 107:030402, 2011.

\bibitem{Amao12}
Elias Amselem, Lars~Eirik Danielsen, Antonio~J. L\'opez-Tarrida, Jos\'e~R.
  Portillo, Mohamed Bourennane, and Ad\'an Cabello.
\newblock Experimental fully contextual correlations.
\newblock {\em Phys. Rev. Lett.}, 108:200405, 2012.

\bibitem{KrRK12}
P.~Kurzy\'nski, R.~Ramanathan, and D.~Kaszlikowski.
\newblock Entropic test of quantum contextuality.
\newblock {\em Phys. Rev. Lett.}, 109:020404, 2012.

\bibitem{Zuao12}
C.~Zu, Y.-X. Wang, D.-L. Deng, X.-Y. Chang, K.~Liu, P.-Y. Hou, H.-X. Yang, and
  L.-M. Duan.
\newblock State-independent experimental test of quantum contextuality in an
  indivisible system.
\newblock {\em Phys. Rev. Lett.}, 109:150401, 2012.

\bibitem{Klao12}
Matthias Kleinmann, Costantino Budroni, Jan-\AA ke~Larsson, Otfried G{\"u}hne,
  and Ad\'an Cabello.
\newblock Optimal inequalities for state-independent contextuality.
\newblock {\em Phys. Rev. Lett.}, 109:250402, 2012.

\bibitem{Cbll13}
Ad\'an Cabello.
\newblock Simple explanation of the quantum violation of a fundamental
  inequality.
\newblock {\em Phys. Rev. Lett.}, 110:060402, 2013.

\bibitem{Amao13}
E.~Amselem, M.~Bourennane, C.~Budroni, A.~Cabello, O.~G{\"u}hne, M.~Kleinmann,
  J.-\AA~. Larsson, and M.~Wie\'sniak.
\newblock Comment on ``{S}tate-{I}ndependent {E}xperimental {T}est of {Q}uantum
  {C}ontextuality in an {I}ndivisible {S}ystem''.
\newblock {\em Phys. Rev. Lett.}, 110:078901, 2013.

\bibitem{Zuao13}
C.~Zu, Y.-X. Wang, D.-L. Deng, X.-Y. Chang, K.~Liu, P.-Y. Hou, H.-X. Yang, and
  L.-M. Duan.
\newblock Reply.
\newblock {\em Phys. Rev. Lett.}, 110:078902, 2013.

\bibitem{TnYO13}
Weidong Tang, Sixia Yu, and C.~H. Oh.
\newblock Greenberger-{H}orne-{Z}eilinger paradoxes from qudit graph states.
\newblock {\em Phys. Rev. Lett.}, 110:100403, 2013.

\bibitem{Spkk05}
R.~W. Spekkens.
\newblock Contextuality for preparations, transformations, and unsharp
  measurements.
\newblock {\em Phys. Rev. A}, 71:052108, 2005.

\bibitem{Appl05b}
D.~M. Appleby.
\newblock The {B}ell-{K}ochen-{S}pecker theorem.
\newblock {\em Stud. Hist. Phil. Mod. Phys.}, 36:1--28, 2005.

\bibitem{Gnvs05}
Marco Genovese.
\newblock Research on hidden variable theories: {A} review of recent
  progresses.
\newblock {\em Phys. Rep.}, 413:319--396, 2005.

\bibitem{Cbll11}
Ad\'an Cabello.
\newblock Quantum physics: {C}orrelations without parts.
\newblock {\em Nature}, 474:456--458, 2011.

\bibitem{Lpao11}
Radek Lapkiewicz, Peizhe Li, Christoph Schaeff, Nathan~K. Langford, Sven
  Ramelow, Marcin Wie\'sniak1, and Anton Zeilinger.
\newblock Experimental non-classicality of an indivisible quantum system.
\newblock {\em Nature}, 474:490--493, 2011.

\bibitem{LnSW11}
Yeong-Cherng Liang, Robert~W. Spekkens, and Howard~M. Wiseman.
\newblock Specker's parable of the overprotective seer: {A} road to
  contextuality, nonlocality and complementarity.
\newblock {\em Phys. Reports}, 506:1--39, 2011.

\bibitem{BFGO11}
Fabio Benatti, Roberto Floreanini, Marco Genovese, and Stefano Olivares.
\newblock Quantum contextuality in {N}-boson systems.
\newblock {\em Phys. Rev. A}, 84:034102, 2011.

\bibitem{Hsgw12}
Yuji Hasegawa.
\newblock Entanglement between degrees of freedom in a single-particle system
  revealed in neutron interferometry.
\newblock {\em Found. Phys.}, 42:29--45, 2012.

\bibitem{Cbll12}
Ad\'an Cabello.
\newblock The role of bounded memory in the foundations of quantum mechanics.
\newblock {\em Found. Phys.}, 42:68--79, 2012.

\bibitem{Cbao12}
Ad\'an Cabello, Elias Amselem, Kate Blanchfield, Mohamed Bourennane, and
  Ingemar Bengtsson.
\newblock Proposed experiments of qutrit state-independent contextuality and
  two-qutrit contextuality-based nonlocality.
\newblock {\em Phys. Rev. A}, 85:032108, 2012.

\bibitem{AbHr12}
Samson Abramsky and Lucien Hardy.
\newblock Logical {B}ell inequalities.
\newblock {\em Phys. Rev. A}, 85:062114, 2012.

\bibitem{KrKs12}
P.~Kurzy\'nski and D.~Kaszlikowski.
\newblock Contextuality of almost all qutrit states can be revealed with nine
  observables.
\newblock {\em Phys. Rev. A}, 86:042125, 2012.

\bibitem{Cbll13b}
Ad\'an Cabello.
\newblock Twin inequality for fully contextual quantum correlations.
\newblock {\em Phys. Rev. A}, 87:010104, 2013.

\bibitem{Chao13}
Jing-Ling Chen, Hong-Yi Su, Chunfeng Wu, Dong-Ling Deng, Ad\'an Cabello, L.~C.
  Kwek, and C.~H. Oh.
\newblock Quantum contextuality for a relativistic spin-1/2 particle.
\newblock {\em Phys. Rev. A}, 87:022109, 2013.

\bibitem{Mrmn93}
N.~David Mermin.
\newblock Hidden variables and the two theorems of {J}ohn {B}ell.
\newblock {\em Rev. Mod. Phys.}, 65:803--815, 1993.

\bibitem{Prs93}
Asher Peres.
\newblock {\em Quantum Theory: Concepts and Methods}.
\newblock Kluwer Academic Publishers, Dordrecht, The Netherlands, 1993.

\bibitem{Hld06}
Carsten Held.
\newblock The {K}ochen-{S}pecker {T}heorem.
\newblock {\em Stanford Encyclopedia of Philosophy},
  http://plato.stanford.edu/entries/kochen-specker/, 2006.

\bibitem{Hrmn11}
Ronnie Hermens.
\newblock The problem of contextuality and the impossibility of experimental
  metaphysics thereof.
\newblock {\em Stud. Hist. Phil. Mod. Phys.}, 42:214--225, 2011.

\bibitem{Bll90}
J.~S. Bell.
\newblock Against measurement.
\newblock In Arthur~I. Miller, editor, {\em Sixty-Two Years of Uncertainty},
  pages 17--31. Plenum Press, New York, 1990.

\bibitem{Grff13}
Robert~B. Griffiths.
\newblock A consistent quantum ontology.
\newblock {\em Stud. Hist. Phil. Mod. Phys.}, 44:93--114, 2013.
\newblock arXiv:1105.3932.

\bibitem{Grff11b}
Robert~B. Griffiths.
\newblock E{P}{R}, {B}ell, and {Q}uantum {L}ocality.
\newblock {\em Am. J. Phys.}, 79:954--965, 2011.
\newblock arXiv:1007.4281.

\bibitem{Hhnb10}
Pierre~C. Hohenberg.
\newblock An introduction to consistent quantum theory.
\newblock {\em Rev. Mod. Phys.}, 82:2835--2844, 2010.
\newblock arXiv:0909.2359.

\bibitem{Grff09b}
Robert~B. Griffiths.
\newblock Consistent {H}istories.
\newblock In Daniel Greenberger, Klaus Hentschel, and Friedel Weinert, editors,
  {\em Compendium of Quantum Physics}, pages 117--122. Springer-Verlag, Berlin,
  2009.

\bibitem{Grff13b}
Robert~B. Griffiths.
\newblock Hilbert space quantum mechanics is noncontextual.
\newblock {\em Stud. Hist. Phil. Mod. Phys.}, To appear, 2013.
\newblock arXiv:1201.1510.

\bibitem{Brn26}
Max Born.
\newblock Zur {Q}uantenmechanik der {S}to\ss vorg{\"a}nge.
\newblock {\em Z. Phys.}, 37:863--867, 1926.

\bibitem{Lloe01}
F.~Lalo{\"e}.
\newblock Do we really understand quantum mechanics? {S}trange correlations,
  paradoxes, and theorems.
\newblock {\em Am. J. Phys.}, 69:655--701, 2001.

\bibitem{Lloe12}
Franck Lalo{\"e}.
\newblock {\em Do We Really Understand Quantum Mechanics?}
\newblock Cambridge University Press, Cambridge, U. K., 2012.

\bibitem{BcVl09}
Guido Bacciagaluppi and Antony Valentini.
\newblock {\em Quantum Theory at the Crossroads}.
\newblock Cambridge University Press, New York, 2009.
\newblock quant-ph/0609184.

\bibitem{Glds12}
Sheldon Goldstein.
\newblock Bohmian mechanics.
\newblock {\em Stanford Encyclopedia of Philosophy},
  http://plato.stanford.edu/entries/qm-bohm/, 2012.

\bibitem{Grff11}
Robert~B. Griffiths.
\newblock Quantum locality.
\newblock {\em Found. Phys.}, 41:705--733, 2011.
\newblock arXiv:0908.2914.

\bibitem{Grff00b}
Robert~B. Griffiths.
\newblock Consistent quantum realism: {A} reply to {B}assi and {G}hirardi.
\newblock {\em J. Stat. Phys.}, 99:1409--1425, 2000.

\bibitem{Bll87}
J.~S. Bell.
\newblock {\em Speakable and Unspeakable in Quantum Mechanics}.
\newblock Cambridge University Press, Cambridge, 1987.

\end{thebibliography}
\end{document}